\documentclass[onecolumn]{article}
\pdfoutput=1
\usepackage[utf8]{inputenc}
\usepackage{amsmath}
\usepackage{enumitem}
\usepackage{array}
\usepackage{tabu}
\usepackage{multirow}
\usepackage{booktabs}
\usepackage{makecell}
\usepackage[ruled]{algorithm2e}
\usepackage{graphicx,xcolor}
\usepackage{float}
\usepackage{color}
\usepackage{pifont}
\usepackage{url}
\usepackage{soul}
\usepackage{authblk}
\usepackage{CJK}
\usepackage{hyperref}
\hypersetup{colorlinks,allcolors=red}
\newcommand{\NParagraph}[1]{{\flushleft\textbf{#1}}} 
\newcommand{\Paragraph}[1]{{\vspace{-1mm}\flushleft\textbf{#1}}} 
\newcommand{\MParagraph}[1]{{\textbf{#1}}} 

\title{Frequency-Aware Physics-Inspired Degradation Model for Real-World Image Super-Resolution}

\author{
    Zhenxing Dong\textsuperscript{1},
    Hong Cao\textsuperscript{1},
    Wang Shen\textsuperscript{2},
    Yu Gan\textsuperscript{3},
    Yuye Ling\textsuperscript{1*},\\
    Guangtao Zhai\textsuperscript{2},
    Yikai Su\textsuperscript{4}
}

\affil{
    \textsuperscript{\rm 1}John Hopcroft Center for Computer Science, Shanghai Jiao Tong University\\
    
    \textsuperscript{\rm 2}Institute of Image Communication and Network Engineering,Shanghai Jiao Tong University\\

    \textsuperscript{\rm 3}Department of Electrical and Computer Engineering, University of Alabama, Tuscaloosa, AL 35401 United States\\

    \textsuperscript{\rm 4}State Key Lab of Advanced Optical Communication Systems and Networks, Shanghai Jiao Tong University\\

    \textsuperscript{\rm *}yuye.ling@sjtu.edu.cn\\

}

\usepackage{bibentry}

\begin{document}
\maketitle

\begin{abstract}

Current learning-based single image super-resolution (SISR) algorithms underperform on real data due to the deviation in the assumed degradation process from that in the real-world scenario.
Conventional degradation processes consider applying blur, noise, and downsampling (typically bicubic downsampling) on high-resolution (HR) images to synthesize low-resolution (LR) counterparts.
However, few works on degradation modelling have taken the physical aspects of the optical imaging system into consideration.
In this paper, we analyze the imaging system optically and exploit the characteristics of the real-world LR-HR pairs in the spatial frequency domain.
We formulate a real-world physics-inspired degradation model by considering both \textit{optics} and \textit{sensor} degradation; The physical degradation of an imaging system is modelled as a low-pass filter, whose cut-off frequency is dictated by the object distance, the focal length of the lens, and the pixel size of the image sensor.
%
%
%
In particular, we propose to use a convolutional neural network (CNN) to learn the cutoff frequency of real-world degradation process. 
The learned network is then applied to synthesize LR images from unpaired HR images.
The synthetic HR-LR image pairs are later used to train an SISR network.
We evaluate the effectiveness and generalization capability of the proposed degradation model on real-world images captured by different imaging systems.
Experimental results showcase that the SISR network trained by using our synthetic data performs favorably against the network using the traditional degradation model.
Moreover, our results are comparable to that obtained by the same network trained by using real-world LR-HR pairs, which are challenging to obtain in real scenes.
%

%
\end{abstract}

\section{Introduction}
\begin{figure}[!t]

\centering\includegraphics[width=1\linewidth]{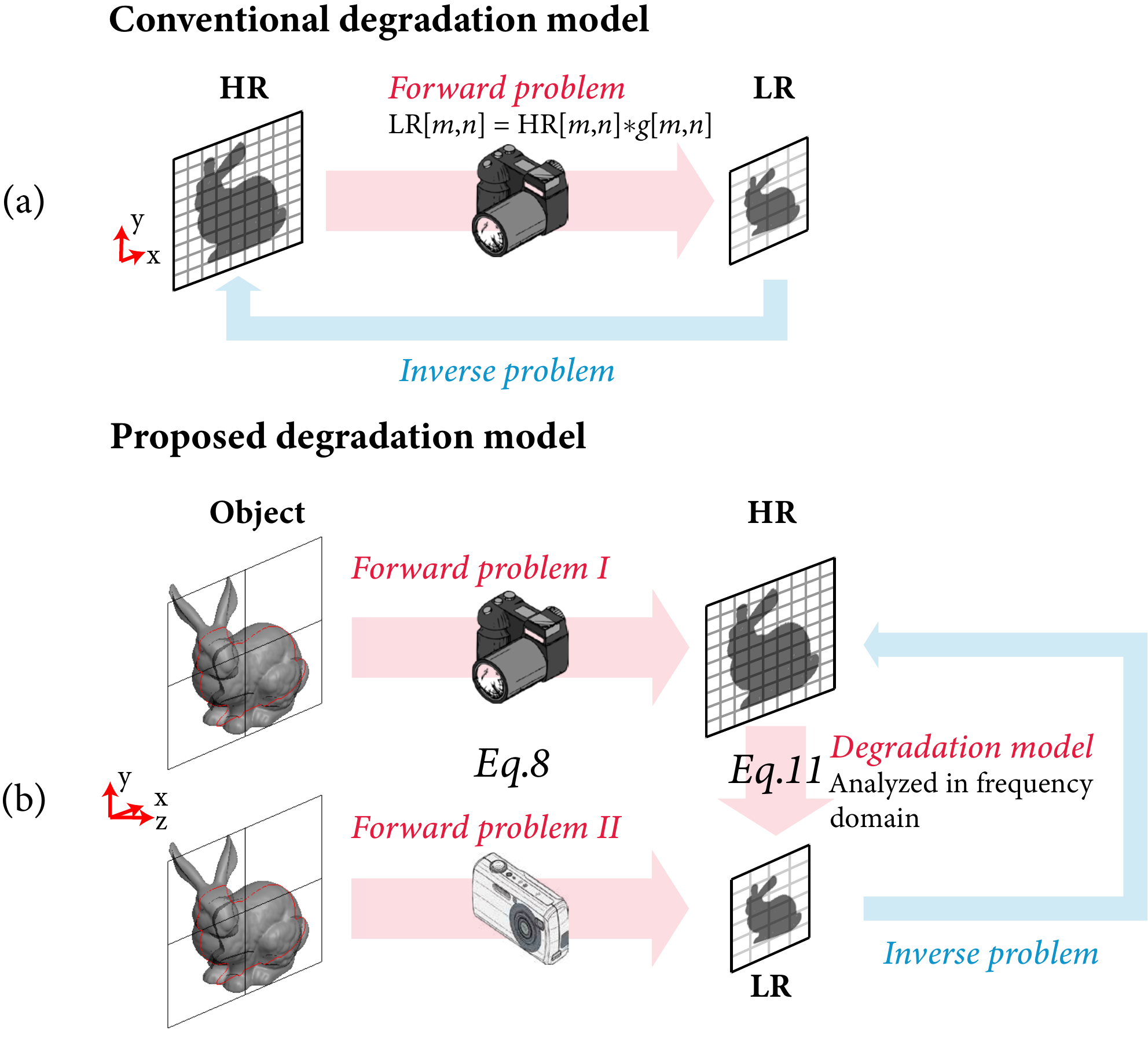}
\caption{Illustration of the conventional and our degradation modes.
We show conventional degradation model, including blur, noise and downsampling (denoted by  $g[m,n]$) in (a), which only considers image-to-image (i.e., HR-to-LR) degradation.However, for the camera acquisition of LR, the object should not be simply regarded as the corresponding HR.Our proposed model considers that HR and LR are both degraded images of a real-world 3D object captured by different devices in (b).Our novel degradation model (denoted by Eq. \ref{eq:eq11}) is analyzed in the frequency domain.Our goal is to solve the inverse problem, which is to recover an HR image obtained by the better device from its corresponding LR recording.}

\label{fig:inverse_prob1}
\end{figure}

With the rise of digital imaging and digital photography in the past decades, persistent endeavors have been devoted to researches that attempt to distill high-spatial-resolution contents from low-spatial-resolution measurements. 
Image super-resolution is a prominent example, in which a high-resolution (HR) image is recovered from a low-resolution (LR) {counterpart}. 
Image super resolution can be categorized as  single-image super resolution (SISR) \cite{glasner2009super} and multi-frame super resolution based on the input data formats; the latter requires a series of slightly misaligned LR images to provide extra {temporal} information, while the former only needs one LR measurement.

Considering that multiple LR images might not be accessible in practical settings, SISR thus becomes an attractive research topic in various fields including medical imaging \cite{10.1093/comjnl/bxm075}, small target detection \cite{Bai_2018_ECCV}, and remote sensing \cite{YUAN2020111716}.

\par Unlike its optical counterparts such as structured illumination microscopy (SIM) \cite{Saxena:15}, localization microscopy \cite{HESS20064258}, and stimulated emission depletion (STED) microscopy \cite{Hell:94}, SISR features a pure algorithmic effort which requires little hardware modifications but fully leverages advanced signal processing techniques.
Conventionally, SISR was performed by traditional methods such as neighborhood embedding \cite{1315043} and dictionary learning \cite{4587647}. With the recent surge in deep learning (DL), SISR algorithms based on the neural networks have been put in the spotlight and their performances have been constantly improved via better network architecture design \cite{dong2014learning,lim2017enhanced,zhang2018image} and refined training strategies \cite{johnson2016perceptual,ledig2017photo}.

\par Despite of the great successes made by the neural network-based SISR algorithms in various public datasets, their capability under real-world circumstances has long been under question \cite{cornillere2019blind}. 
It is because most of the existing SISR networks are trained and evaluated on synthetic datasets, in which the LR images are not physically but numerically acquired by applying degradation models including Gaussian blur, Bicubic downsampling, and Gaussian noise on the corresponding HR images in the spatial domain, as shown in Fig.\ref{fig:inverse_prob1}(a).
%

Nonetheless, the inherent mappings of the real-world HR-LR image pairs could be far more sophisticated than the aforementioned analytical models.
To bridge this gap, researchers \cite{chen2019camera,zhang2019zoom} attempt to train the SISR neural networks by using HR-LR image pairs that are collected in the real world, so that the trained networks could implicitly learn the real-world degradation model and get its generalization ability enhanced. 
Unfortunately, preparing such a database is very challenging; excessive labours, specially designed hardware, and careful postprocessings are often required to construct a small-scale dataset. For example, RealSR \cite{cai2019toward} only contains 559 scenes captured by two different devices.

\par In this paper, we propose a physics-inspired degradation model for real-world SISR tasks.
We suggest that the conventional degradation model, which is illustrated in Fig. \ref{fig:inverse_prob1}(a), might not be accurate: for the camera acquisition of LR, the object should not be simply regarded as the corresponding HR.
We instead argue that HR and LR are both degraded images of a real-world 3D object as shown in Fig. \ref{fig:inverse_prob1}(b), in which, for example, the HR is acquired by a DSLR camera and the LR is captured by a cellphone camera.
We further elucidate that the goal of a real-world SISR task is to recover an HR image obtained by a better device from its corresponding LR recording.
To validate our hypothesis, we study the optical imaging systems in Gaussian optics' regime, analytically derive the degradation model, which could be viewed as a \textit{low-pass} transfer function, between HR and LR in the spatial frequency domain, and experimentally show that the {characteristics} of the transfer function is jointly determined by the object distance, the focal length of the lens, and the pixel size of the image sensor (see Fig. \ref{fig:degra_func1}).

\par Once the degradation model is properly defined, we move on to solve the inverse problem. 
We propose to use a convolutional neural network to learn the SISR degradation in the spatial frequency domain.
It is worth noting that the network is supposed to be aware of the object distance as well as the camera specifications via the supervised learning of HR image's spectrum and the corresponding low-pass filter's \textit{cutoff} bandwidth.
The learned prediction network can thus be utilized to synthesize \textit{real-world-like} LR images from arbitrary HR images.
The thus generated HR-LR pairs could be further used to train SISR networks.
Experimental results (Fig. \ref{fig:RealSR1}) show that the SISR network trained by our method obtain the best performance.

The contributions of this paper are summarized below:

\begin{itemize}[nosep, leftmargin=0.45cm]
\item We propose a novel physics-inspired degradation model to incorporate both optics and sensor degradation in the spatial frequency domain for real-world SISR problems.
\item We suggest the abovementioned degradation model could be implicitly learned by a neural network and later applied on arbitrary HR images to synthesize their real-world-\emph{like} LR counterparts.
\item We demonstrate the generalization and effectiveness of our degradation model in various imaging systems for real-world image super-resolution.
\end{itemize}

\section{Related Work}

\NParagraph{Degradation Model:}
As mentioned in the previous section, SISR networks trained by traditionally synthesized HR-LR pairs might not necessarily learn the actual degradation process. Some works \cite{9022593, Ji_2020_CVPR_Workshops, zhang2021designing} attempted to approximate the real-world degradation model by complicating the blur, downsampling, and noise models. While impressive results were obtained in this way, the resultant degradation models often lack intrinsic interpretability capability due to the absence of physics during the model constructions. Moreover, their proposed methods only work on the same datasets as their degradation models. 
\Paragraph{Blind SISR Networks:} 
The research community has also spent a lot of efforts on blind super-resolution, in which key parameters of the degradation process are assumed unknown. We can roughly divide the blind SISR related works into two genres:
\par One approach is to directly estimate the unknown blur kernel. Gu et al. \cite{gu2019blind} propose an Iterative Kernel Correction
(IKC) architecture, which uses the intermediate SR results to iteratively correct the estimated blur kernel. Zhou et al. \cite{zhou2019kernel} study a real-world super-resolution dataset (DSLR) \cite{ignatov2017dslr} and propose a Kernel Modeling Super-Resolution (KMSR) network.
While the performance of the network is improved compared with other works in real world, and the estimated kernel lacks a physical interpretation and does not perform well if being generalized to other imaging devices.

\par The other way is to adopt a data-driven methodology: some researchers also seek to construct real-world datasets to implicitly infer the underlying degradation model. 
Zhang et al. \cite{zhang2019zoom} collect 500 HR-LR image pairs at multiple focal lengths by using Sony cameras (SR-RGB). Unfortunately, the image pairs are not registered, so that the SR-RGB dataset cannot be directly used to train SISR networks. Cai et al. \cite{cai2019toward} capture HR and LR data pairs in multiple real-world scenes at multiple focal lengths, and propose a new pairing algorithm to align the data pairs. However, capturing the real-world SISR dataset requires a lot of manpower and material resources, and the size of the dataset is also limited. 

\section{Frequency-Aware Physics-Inspired Degradation Model}
In the following section, we propose a frequency-aware physics-inspired degradation model that could characterize both \textit{optics} and \textit{sensor} degradation processes for digital photography. The model is derived on the basis of linear system theory and Gaussian optics \cite{smith2008modern}. 
%

\begin{figure}[H]
\centering\includegraphics[width=0.8\textwidth]{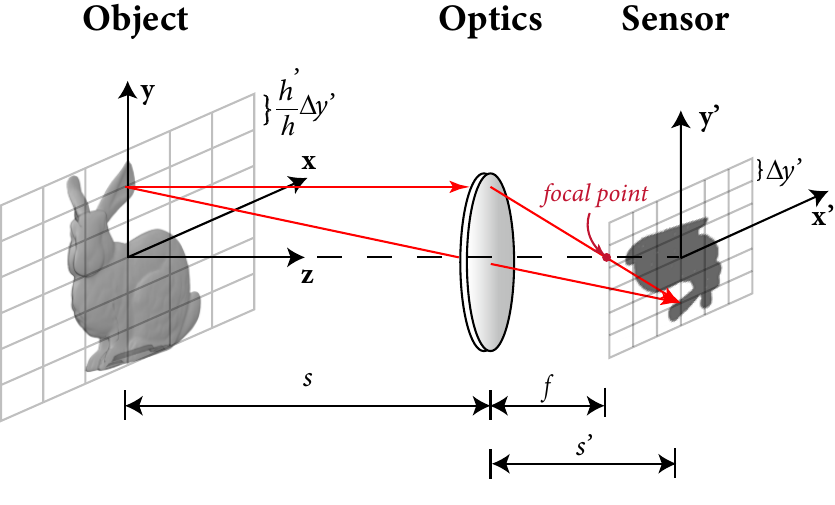}
\vspace{-10pt}
\caption{{Thin lens imaging process}. We denote focal length, object distance, object height, image distance, imaging height, and pixel size by 
$f$, $s$, $h$, $s'$, $h'$, and $\Delta y'$.}
\label{fig:thin_img1}
\vspace{-5pt}
\end{figure}

\Paragraph{Image formation:}
We focus our discussion on the optics of consumer-level cameras, e.g., digital single-lens reflex (DSLR) cameras and cellphone cameras. 
Optical model of those cameras can be approximately considered as a thin lens~\cite{smith2008modern}.
We depicted the image formation in Fig. \ref{fig:thin_img1}, and the corresponding \textit{thin lens formulae} are given by:
\begin{equation}
\left\{
\begin{aligned}
\frac{1}{f} & = \frac{1}{s'} + \frac{1}{s}, \\
\frac{h}{h'} & = \frac{s}{s'},
\label{eq:1}
\end{aligned}
\right.
\end{equation}
where $f$ is the focal length of the equivalent thin lens, $s$ is the object distance, $s'$ is the image distance, $h$ is the object height, and $h'$ is the image height. Considering that  the lens' focal length $f$ is usually much smaller than the object distance $s$ in photography except for macro photography, we could safely omit its contribution and obtain:
\begin{equation}
f \approx s'.
\label{eq:2}
\end{equation}

By plugging Eq. \ref{eq:2} back to the second line of Eq. \ref{eq:1}, we could obtain the following equation with regard to the magnification $M$ between the image and object:
\begin{equation}
M = \frac{h'}{h} = \frac{{{f}}}{s}
\label{eq:3}
\end{equation}

\Paragraph{Optics degradation model:}
Despite of the demagnification, the formed image is also distorted by the optical aberrations of the image system. This effect can be fully described by a point spread function (PSF) $p_{s}(x,y)$ in the context of linear system theory \cite{goodman2005introduction}. It is should be noted that $p_{s}(x,y)$ is shift-invariant in the transverse directions ($XY$ plane) but has a strong axial dependency on the object distance $s$. Therefore, the optically degraded image $I_o(x,y)$ could be given by:
\begin{equation}
{I_o(x,y)} = o(x,y) * p_{s}(x,y),
\label{eq:4}
\end{equation}
where the originally three dimensional object $o(x,y,z)$ is now replaced by a flat projection $o(x,y)$, and $*$ is the convolution operation. This approximation is largely accurate if the object distance $s$ is much larger than the axial extent of the object.

\Paragraph{Sensor degradation model:}
We further take the sensor into account. Without loss of generality, we assume the pixel size of the sensor ($\Delta x'$ and $\Delta y'$) has an aspect ratio of 1, i.e. $\Delta x' = \Delta y'$. Therefore, the maximum spatial frequency $\xi'_{\textrm{sensor}}$ could be detected by the sensor in the image plane is given by Nyquist–Shannon sampling theorem:
\begin{equation}
    \xi'_{\textrm{sensor}}=\frac{1}{2\Delta x'}=\frac{1}{2\Delta y'}.
\label{eq:5}
\end{equation}
\par Consider the scaling relationship between the object plane and image plane as shown in Fig. \ref{fig:thin_img1}, we could obtain the corresponding spatial sampling interval $\Delta x$ and spatial sampling frequency $\xi_{\textrm{object}}$ in the object plane:
\begin{equation}
\left\{
\begin{aligned}
    \Delta x &= {M}\cdot\Delta x'=\frac{f}{s}\cdot\Delta x', \\
    \xi_{\textrm{object}}&={M}\cdot \xi'_{\textrm{sensor}}=\frac{f}{s}\cdot\frac{1}{2\Delta x'}.
\end{aligned}
\right.
\label{eq:6}
\end{equation}
It is important to recognize that the spatial sampling interval $\Delta x$ is determined not only by the sensor pixel size $\Delta x'$ and the camera's focal length $f$ as described in previous works \cite{10.1145/3197517.3201333} but also the object distance $s$. 

.
\Paragraph{Frequency-aware physics-inspired degradation model:} The complete imaging model could be constructed by combining the results mentioned above:
\begin{equation}
\centering
\hspace{-0.5mm}
\begin{aligned}
I(m\Delta x', n\Delta y') &=\mathcal{S}\left[I_{o}(x, y)\right]+\eta \\
&=\mathcal{S}[o(x, y) * p_{s}(x,y)]+\eta \\
&=o(x, y) * p_{s}(m\Delta x, n\Delta y) + \eta \\
&=o(x, y) \!*\! p_{s}(\frac{f}{s}m\Delta x',\! \frac{f}{s}n\Delta y') \!+\! \eta,
\end{aligned}
\label{eq:7}
\end{equation}
where $I$ is the sensor image, $\mathcal{S}$ represents for the sampling operation, $m$ and $n$ are the sampling indices, and $\eta \sim \mathcal{N}(0,{\sigma ^2})$ denotes the Gaussian read noise. Properties \cite{zhou2019kernel} regarding to the $S$ operation are utilized to simplify the equation.

%
We can further derive the explicit expressions for HR and LR images of the same object acquired by two different imaging systems as follows:

\begin{equation}
\centering
\hspace{-1.5mm}
\left\{
\begin{aligned}
\text{HR}(m_1\Delta x_1',\! n_1\Delta y_1')  \!=\! o(x,y) 
 \!*\! p_{1,s}(\frac{f_1}{s}m_1\Delta x_1',\! \frac{f_1}{s}n_1\Delta y_1')\!+\!\eta_1,
\\
\text{LR}(m_2\Delta x_2',\! n_2\Delta y_2') \!=\! o(x,y) \!*\! p_{2,s}(\frac{f_2}{s}m_2\Delta x_2',\! \frac{f_2}{s}n_2\Delta y_2')\!+\!\eta_2. 
\label{eq:8}
\end{aligned}
\right.
\end{equation}

It is obvious that HR and LR images are in general of distinct sizes. 
Moreover, the field-of-views (FOVs) are also different, which makes cropping, upsampling, and registration mandatory before analyzing its degradation \cite{cai2019toward}.

\par Without loss of generality, we could assume the sensors used in two systems are identical ($m_1=m_2$, $n_1=n_2$, $\Delta x_1'=\Delta x_2'=\Delta y_1'=\Delta y_2'$) and the ratio between the effective focal length is $2:1$. In this case, the central part of the LR image need to be cropped out and upsampled to match the FOV of the whole HR image. Once this is done, we can transfer the matched HR-LR image pair to the spatial frequency domain: 

\begin{equation}
\centering
\hspace{-1.5mm}
\left\{
\begin{aligned}
\mathcal{F}(\textrm{HR}(m\Delta x', n\Delta y'))\!&=\!\mathcal{O}(f_x, \!f_y) \! \cdot \! {\mathcal{P}_{1,s}}(m\Delta f_x',\! n\Delta f_y') \!+\! \mathcal{F}(\eta_1)
,\\
\mathcal{F}(\textrm{LR}(m\Delta x', n\Delta y'))\!&=\!\mathcal{O}(f_x, \!f_y) \! \cdot \! {\mathcal{P}_{2,s}}(m\Delta f_x',\! n\Delta f_y') \!+\! \mathcal{F}(\eta_2)
,
\label{eq:9}
\end{aligned}
\right.
\end{equation}

where $\mathcal{F}$ denotes 2-D Fourier transform (FT), $f_{x,y}$ are the spatial frequency in $x$ and $y$ direction, $\mathcal{O}$ and $\mathcal{P}_{1,2}$ represents for the 2-D FT of $o(x,y)$ and $p_{1,2}(x,y)$, respectively.
The transfer function $\mathcal{G}$ of the proposed degradation model can be retrieved by simply dividing the spectra of HR image by that of the LR image:
\begin{equation}
\begin{aligned}
\mathcal{G}_{s}(m\Delta f_x', n\Delta f_y') &=\frac{\mathcal{F}(\textrm{LR}(m\Delta x', n\Delta y'))}{\mathcal{F}(\textrm{HR}(m\Delta x', n\Delta y'))} \\
&\approx\frac{\mathcal{P}_{{2,s}}(m\Delta f_x', n\Delta f_y')}{\mathcal{P}_{{1,s}}(m\Delta f_x', n\Delta f_y')},
\end{aligned}
\label{eq:10}
\end{equation}
where we omit $\mathcal{F}(\eta_1)$ and $\mathcal{F}(\eta_2)$ based on the assumption that the signal-to-noise-ratio (SNR) of both HR and LR images are relatively high.

Therefore, we can formally write the forward model for the SISR problem as:
\begin{equation}
\centering
\textrm{LR}[m,n]\!=\!\left(\mathcal{F}^{-1}\{\mathcal{F}(\textrm{HR}[m,n])\! \cdot \! \mathcal{G}_{s}[m,n]\}\right){ \downarrow _N}\!+\!\eta,
\label{eq:eq11}
\end{equation}
where $\mathcal{F}^{-1}$ is the inverse 2-D Fourier transform, $\downarrow _N$ is downsampling operation with a factor of $N$.
\par In summary, we \textit{first} systematically and theoretically demonstrate the real-world SISR degradation process based on Gaussian optics, and propose a frequency-aware physics-inspired degradation model, which associated with object distance, focal length and pixel size of the imaging system.




\begin{figure*}[t]
\footnotesize
\centering\includegraphics[width=1\linewidth]{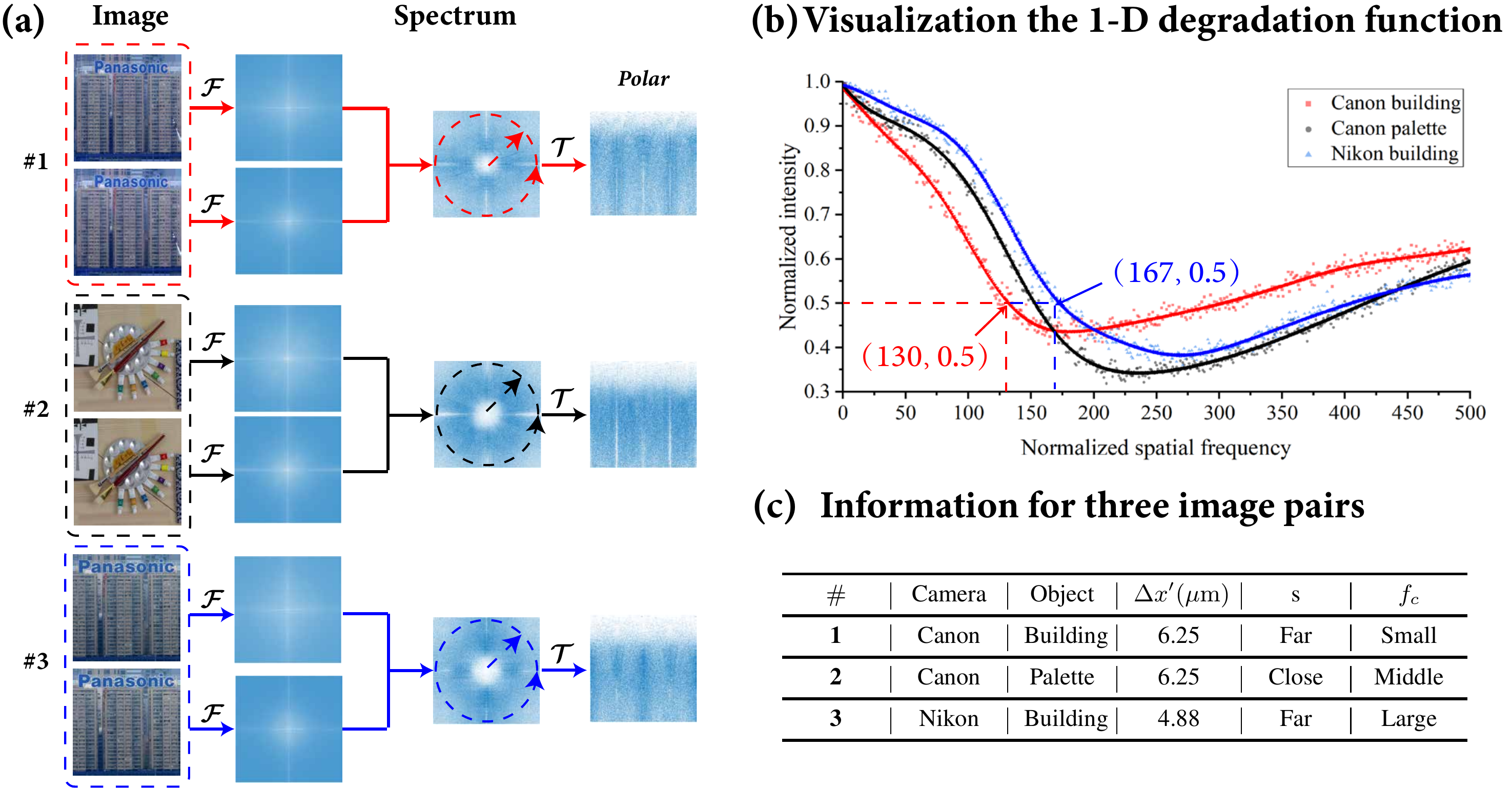}

\vspace{-5pt}
\caption{{Visualization of real-world degradation model}. (a) A series of operations on three HR-LR pairs of RealSR \cite{cai2019toward} in order to extract real degradation information in the frequency domain. (b) Visualization the 1-D degradation function and mark the cut-off frequency points of the degenerate function of Canon building and Nikon building; (c) Some important information for three image pairs and cameras. $\mathcal{F}$ is 2-D Fourier transform and $\mathcal{T}$ is the conversion of Cartesian coordinate  to polar coordinate.}
\label{fig:degra_func1}
\vspace{-10pt}
\end{figure*}

\NParagraph{Interpretability:}
To illustrate the dependencies of the degradation model on the object distance $s$ and sensor pixel size $\Delta x'$, we evaluate the model on a popular real-world SISR dataset RealSR~\cite{cai2019toward}. All image pairs are captured by Canon EOS 5D Mark III and Nikon D810, both of which are equipped with one 24-105mm, f/4.0 zoom lens.

\par We select three pairs of HR-LR images from the dataset.
The HR and LR images are captured using the focal lengths of 105 mm and 55 mm, respectively, and more detailed information is provided in Fig. \ref{fig:degra_func1}(c).
A series of operations are performed on the HR-LR pairs to acquire the corresponding transfer function $\mathcal{G}_s$ as shown in Fig. \ref{fig:degra_func1}(a). Firstly, we perform 2-D FTs on both HR and LR images, and take the logarithm of spectra. We could then obtain the transfer function $\mathcal{G}_s$ by simply subtracting the spectrum of HR from that of LR. After that, the coordinate system is changed from Cartesian to polar to better visualize the isotropic low-pass characteristic of the transfer function $\mathcal{G}_s$. Finally, we take the exponential of $\mathcal{G}_s$ and average it over angular direction. The resultant curves for three HR-LR pairs are plotted in Fig. \ref{fig:degra_func1}(b). 
We have the following two observations:

\MParagraph{(1) The dependency of the degradation model on the object distance $s$:}
One interesting observation could be made by comparing the transfer functions $\mathcal{G}_s$ for HR-LR pair $\#1$ and $\#2$ as shown in Fig. \ref{fig:degra_func1}(b) is that the cutoff frequency for the former is lower than that of the latter. This result matches the prediction made by Eq. \ref{eq:6} exactly, which states that the spatial sampling frequency in the object plane $\xi_{\textrm{object}}$ is inversely proportional to the object distance $s$, if we consider the building of the image pair $\#1$ is located much farther than the palette of pair $\#2$. More similar results could be found in the supplementary materials. 

\MParagraph{(2) The dependency of the degradation model on sensor pixel size $\Delta x'$:} We analyze the influence of pixel size by comparing the results of pair $\#1$ and pair $\#3$, where the same building (i.e. fixed object distance $s$) is captured by distinct cameras:

%
 
\begin{itemize}[nosep, leftmargin=0.45cm]
\item \textit{Object sampling frequency ratio $\alpha$}: The ratio of the object sampling frequency between the two cameras used in the experiment could be theoretically calculated by using published technical specifications:
\begin{equation}
\centering
\alpha =\frac{\xi_{\textrm{object,Nikon}}}{\xi_{\textrm{object,Canon}}}=\frac{\Delta x'_{\textrm{Canon}}}{\Delta x'_{ \textrm{Nikon}}}=\frac{6.25 \,\mu m}{4.88 \,\mu m} \approx 1.28.
\label{eq:12}
\end{equation}

\item \textit{Cutoff frequency ratio $\beta$}: The actual ratio between the cutoff frequencies, on the other hand, could be directly inferred from Fig. \ref{fig:degra_func1}(b):
\begin{equation}
\hspace{-3mm}
\beta = \frac{{{f_{c,\textrm{Nikon}}}}}{{{f_{c,\textrm{Canon}}}}} = \frac{{167}}{{130}} \approx 1.28.
\label{eq:13}
\end{equation}
\end{itemize}

It is clear that the theoretically predicted $\alpha$ matches the experimentally derived $\beta$, which verifies the inverse proportionality between the object sampling frequency $\xi_{\textrm{object}}$ and sensor pixel size $\Delta x'$.

\par In summary, we showcase the physical soundness of the proposed degradation model by conducting a series of controlled experiments. Most importantly, to our best knowledge, this is the \textit{first} time that both object distance $s$ and sensor pixel size $\Delta x'$ are explicitly integrated in a degradation model for SISR.

\section{Proposed Method}

\begin{figure}[H]
\centering\includegraphics[width=0.8\linewidth]{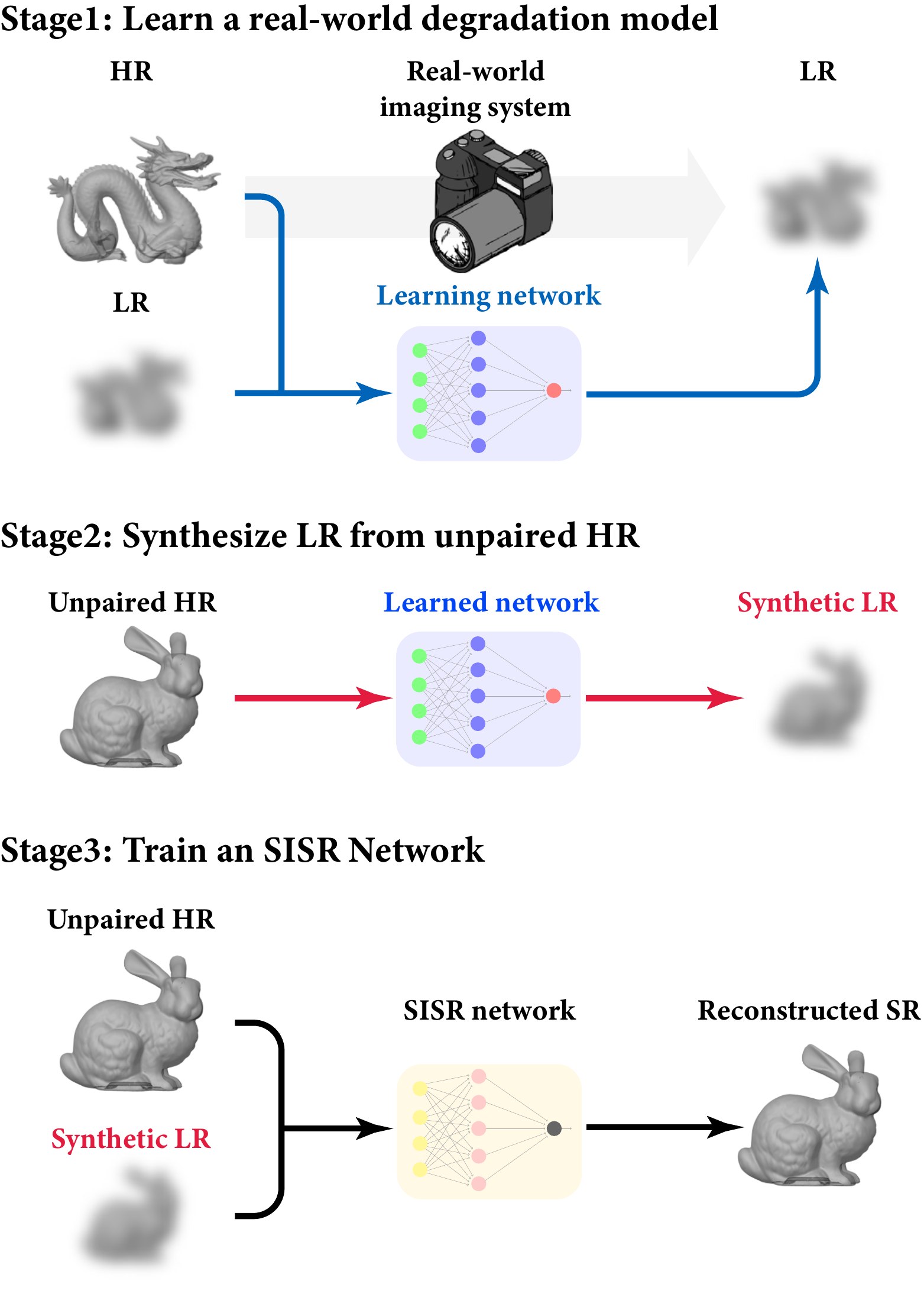}
\vspace{-10pt}
\caption{{Pipeline of our proposed model}. Stage1: Utilize existing real-world SR datasets to learn a real-world imaging system; Stage2: Use the learned imaging system to synthesize proposed LR images from unpaired real HR images; Stage3: Train an SISR network on our proposed dataset.} 
\vspace{-5pt}
\label{fig:framework1}
\end{figure}

The proposed pipeline for real-world SISR is showed in Fig. \ref{fig:framework1}.
It consists of three stages.

\begin{figure}[!t]
\vspace{-5pt}
\centering\includegraphics[width=1\linewidth]{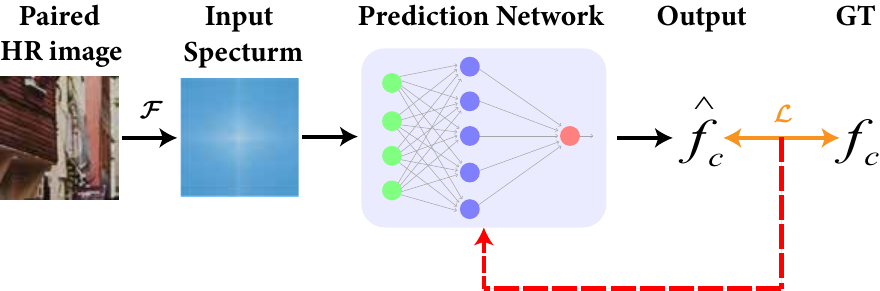}
\caption{Learning the cutoff $f_c$ of a imaging system. Although we present that the paired HR image is an RGB image, the neural network is actually trained in color channel.
}
\vspace{-5pt}
\label{fig:leran_img1}
\end{figure}


\NParagraph{Stage 1: Learn a real-world degradation model:}
%
In the previous section, we have demonstrated that the transfer function $\mathcal{G}_s$ could be largely viewed as a low-pass filter (LPF) in the spatial frequency domain, and the most important parameter for the LPF is the cutoff frequency $f_c$ which is dictated by both $s$ and $\Delta x'$. Therefore, we proposed to simply learn the cutoff frequency $f_c$ instead of the full transfer function $\mathcal{G}_s$.

\par First of all, we divide each image pair $\left\{ {\left. {{I^{\textrm{HR}}},{I^{\textrm{LR}}}} \right\}} \right.$ that is captured by the same device from RealSR \cite{cai2019toward} into $J$ pairs of overlapping patches ${\left\{ {\left. {{I_j}^{\textrm{HR}},{I_j}^{\textrm{LR}}} \right\}} \right.}$, where $j = 1,2,...,J$. 
For every pair of patches, we follow the procedure described in Fig. \ref{fig:degra_func1} to obtain the cutoff frequency ${f_{{c_j}}}$. More detailed process can be found in supplementary material Algorithm 1.

\begin{table*}
\footnotesize
\centering
\caption{{Comparison quantitative results of average PSNR and SSIM on Canon and Nikon dataset from RealSR \cite{cai2019toward}, including FSSR-DPED, FSSR-JPEG, RealSR-DPED  , RealSR-JPEG, BSRGAN, Proposed, and Proposed$^\dag$} 
%
 The numbers in {\color{red}red} and {\color{blue}blue} represent the best and second-best results.
 }\label{tab:tab2}
\vspace{5pt}
\scalebox{0.7}{
\begin{tabular}{m{1.5cm}<{\centering}m{1.5cm}<{\centering}|m{1.5cm}<{\centering}m{1.5cm}<{\centering}m{1.5cm}<{\centering}m{1.5cm}<{\centering}m{1.5cm}<{\centering}|m{1.5cm}<{\centering}m{1.5cm}<{\centering}}
\toprule
{Testing set}            & {Metric}  & {FSSR-DPED} & {FSSR-JPEG} & {RealSR-DPED} & {RealSR-JPEG}  & {BSRGAN}  &{Proposed} &{Proposed$^\dag$}\\
\midrule
\multirow{2}{*}{{Canon}}   & PSNR              & 25.18     & 23.46               &23.08          & {\color{blue}25.67}     & 25.17       & 25.59           & {\color{red}25.69}       \\
                         & SSIM          & {\color{blue}0.835}          & 0.753               & 0.777   & 0.820     & 0.834          & {\color{red}0.847}               & {\color{red}0.847}         \\

\midrule
\multirow{2}{*}{{Nikon}}   & PSNR          & 24.48     & 22.32     & 21.84              & 25.54      & 24.60              & {\color{red} 25.70  }             & {\color{blue}25.58}      \\
                         & SSIM             &0.807     & 0.674               & 0.731   & 0.791     & 0.805       & {\color{red} 0.832  }               & {\color{blue}0.830}            \\
                                            
\bottomrule
\end{tabular}}
\vspace{-5pt}
\end{table*}

\par We then train a prediction network $\mathcal{R}$ to learn the mapping from the HR spectrum $\mathcal{F}(I_j^{\textrm{HR}})$ to the corresponding cutoff frequency ${f_{{c_j}}}$ for a specific device by supervised learning as is depicted in Fig. \ref{fig:leran_img1}. In this study, we choose ResNet34 \cite{he2016deep} as the prediction network.
%
%
\Paragraph{Stage 2: Synthesize LR images from unpaired HR images:}
Once the mapping from an input HR spectrum to cutoff frequency in \textit{source} imaging system is learned, we can use it to generate \textit{real-world-like} from arbitrary HR images for a \textit{target} imaging system. Here, we select SR-RGB \cite{zhang2019zoom} captured with the focal length of 105mm as the source of HR images.

We first preprocess the selected HR images in the same way as laid out in Stage 1: each unpaired HR image $I^{\textrm{HR,unpaired}}$ from SR-RGB is divided into $J$ overlapping patches $I_j^{\textrm{HR,unpaired}}$. We then input $\mathcal{F}(I_j^{\textrm{HR,unpaired}})$ to the trained $\mathcal{R}$ to predict the corresponding cutoff frequency $f_{c_j,\textrm{source}}$. For simplicity, We neglect the spatial dependency of the degradation model within the same image, and use the image-wise average cutoff frequency $f_{c,\textrm{source}} =\frac{1}{J}\sum\limits_j {f_{j_c,\textrm{source}}}$ instead.

After acquiring $f_{c,\textrm{source}}$, a correction is made to handle the misfit between the pixel size of the source system and that of the target system: ratio $\alpha$ is first calculated based on Eq. \ref{eq:12} and later multiplied by $f_{c,\textrm{source}}$ to retrieve the correct $f_{c,\textrm{target}}$.

Finally, a second-order Butterworth low-pass filter $\mathcal{B}$ is constructed to approximate the transfer function $\mathcal{G}_s$, and we apply Eq. \ref{eq:eq11} to synthesize the \textit{real-world-like} LR images from the input unpaired HR images. Algorithm 2 in the supplementary material shows the detailed process.

\Paragraph{Stage3: Train an SISR Network:}
Since the focus of the current study is on the degradation model instead of the SISR network, we utilize an existing SISR network EDSR \cite{lim2017enhanced}. It should be noted that the proposed degradation model does not limit the use of other SISR networks. 


\section{Experiments}

\subsection{Implementation Details}

\NParagraph{Dataset and training details:}
In Stage 1, two distinct prediction networks are used to learn Canon EOS 5D Mark III and Nikon D810 imaging systems based on RealSR dataset \cite{cai2019toward}, respectively. 
The parameters are set the same for both networks: the batch size is 16,  the patch size is $320 \times 320$, the number of epochs is 100, the number of training samples is 12,800, the learning rate is $1 \times {10^{{\rm{ - }}3}}$, and the optimizer is RMSprop (\verb"weight_decay=10^{-8}", \verb"momentum=0.9"). 

\par In Stage 3, a total number of four synthetic datasets are generated, and they can be categorized into two sets (Ours and Ours$^\dag$) based on whether the source system is identical to the target system: 
 
\begin{itemize}[nosep, leftmargin=0.45cm]
\item \textbf{Ours}: 

(1) Use the predicting network learned from a Canon system to synthesize \emph{Canon-like} LR images. 

(2) Use the predicting network learned from a Nikon system to synthesize \emph{Nikon-like} LR images.
\item \textbf{Ours$^\dag$}: 

(3) Use corrected predicting network learned from a Canon system to synthesize \emph{Nikon-like} LR images.

(4) Use corrected predicting network learned from a Nikon system to synthesize \emph{Canon-like} LR images.
\end{itemize}

For the training of EDSR, the patch size is set to $96 \times 96$, the batch size is fixed to 16, the initial value of the learning rate is $1 \times {10^{{\rm{ - }}4}}$, and the number of training epochs is 200. The Adam default parameters (${\beta _1}$ $=$ 0.9, ${\beta _2}$ $=$ 0.999 and $\varepsilon$ $=$ ${10^{{\rm{ - }}8}}$) are used in the process of optimizing the network. 
All experiments are trained using NVIDIA GeForce RTX 3090.

\NParagraph{Evaluation metric:}
Peak Signal-to-Noise Ratio (PSNR) and Structural Similarity Index (SSIM) \cite{1284395} are employed to evaluate the performance of competing methods.

\subsection{Comparison to state-of-the-arts}

\NParagraph{Baselines:}
We compare our network (denoted by Proposed) on RealSR dataset \cite{cai2019toward} with several state-of-the-art methods, including FSSR-DPED \cite{9022593}, FSSR-JPEG \cite{9022593}, RealSR-DPED  \cite{Ji_2020_CVPR_Workshops}, RealSR-JPEG \cite{Ji_2020_CVPR_Workshops}, and BSRGAN \cite{zhang2021designing}.

\NParagraph{Objective results:}
As listed in Table \ref{tab:tab2}, Proposed surpasses BSRGAN by 0.42 dB in PSNR and 0.013 in SSIM, respectively, in the Canon testing set. For the Nikon testing set, Proposed further widens the lead by 1.1 dB in PSNR and 0.027 in SSIM, respectively.

\begin{figure*}[t]
\centering\includegraphics[width=1.0\linewidth]{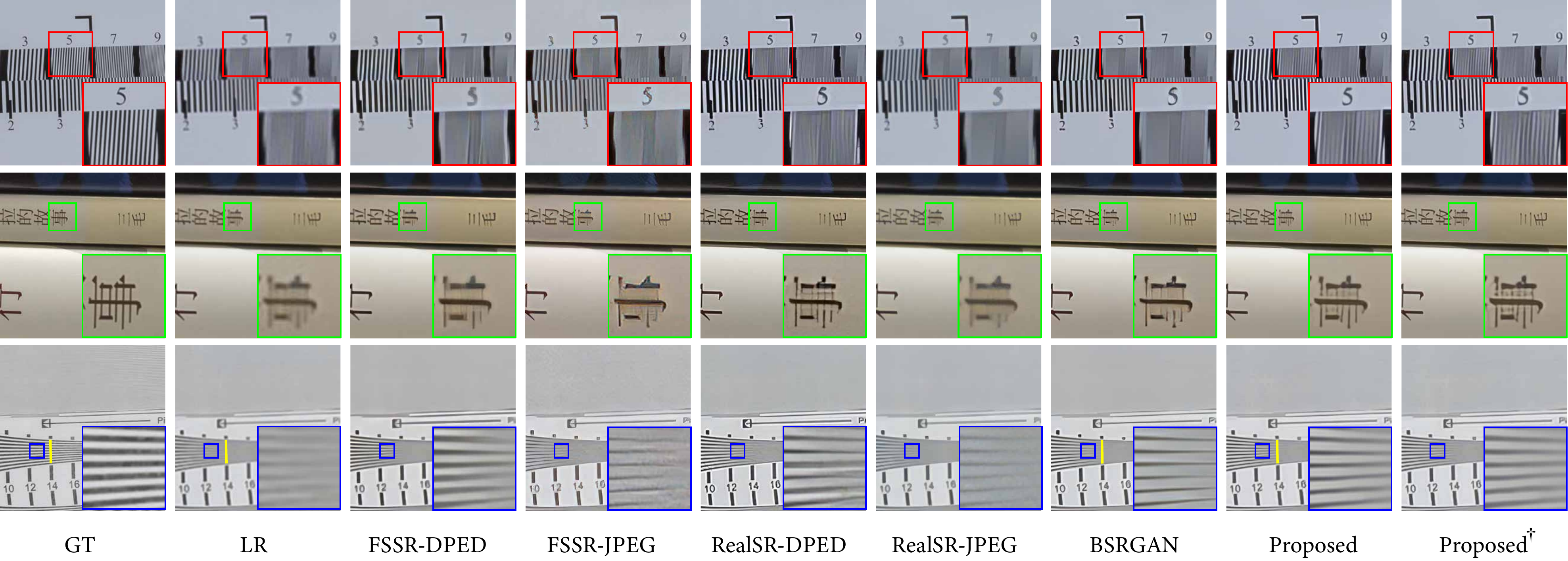}
\vspace{-15pt}
\caption{{Comparison Qualitative $\times4$ SR results of on Canon and Nikon dataset from RealSR \cite{cai2019toward}, including FSSR-DPED, FSSR-JPEG, RealSR-DPED, RealSR-JPEG, BSRGAN, Proposed, and Proposed$^\dag$.}}
\vspace{-5pt}
\label{fig:RealSR1}
\end{figure*}

\NParagraph{Visual results:}
Fig. \ref{fig:RealSR1} shows the visual results obtained by the abovementioned networks. It is clear that network trained by our dataset possesses the best spatial resolution. In particular, Proposed can completely resolve the periodic black and white stripes in both horizontal (first row of Fig. \ref{fig:RealSR1}) and vertical directions (third row of Fig. \ref{fig:RealSR1}), while artifacts are observed in SOTAs.

\begin{figure*}[t]
\centering\includegraphics[width=1\linewidth]{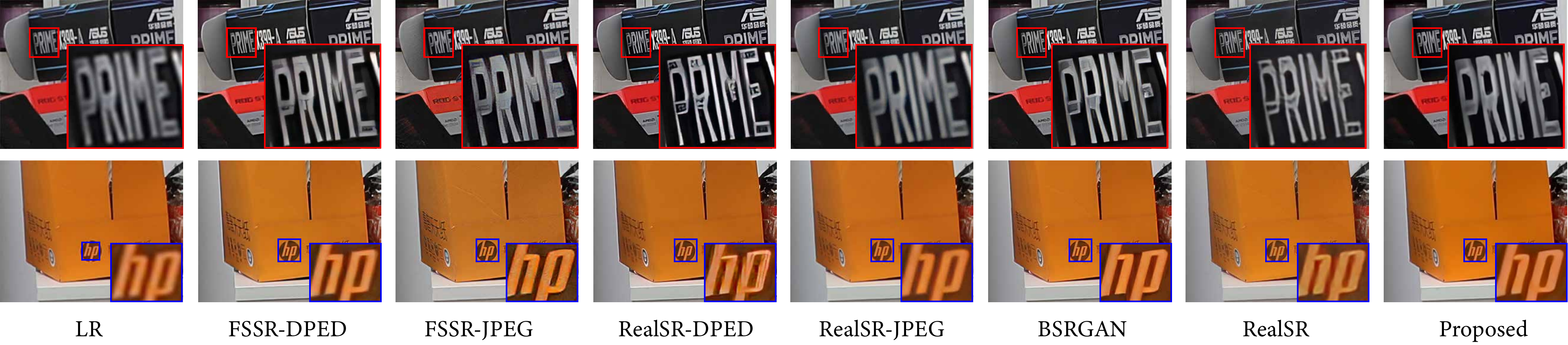}
\vspace{-15pt}
\caption{{Visual $\times4$ SR results of on real-captured images, including FSSR-DPED, FSSR-JPEG, RealSR-DPED, RealSR-JPEG, BSRGAN, EDSR trained on RealSR (denoted by RealSR), and Proposed.}}
\vspace{-5pt}
\label{fig:Real_images1}
\end{figure*}

\NParagraph{Experiments on author-captured images:}
To further confirm that the proposed model can be adapted to real-world SISR, we use another DSLR camera (Canon EOS 5D Mark IV, $\Delta x'=5.36$ $\mu$m, $f=28$ mm) to capture real-world images as testing LRs. The super-resolved results by using different models are all visualized in Fig. \ref{fig:Real_images1}. It is clear that the network trained on our dataset achieves better visual quality than SOTAs with more natural rendering and less artifacts. Although BSRGAN's visual quality results are comparable with Proposed, we can't neglect its much larger training set and superior network.


\subsection{Model Analysis}

\begin{figure}[H]
\centering\includegraphics[width=0.7\linewidth]{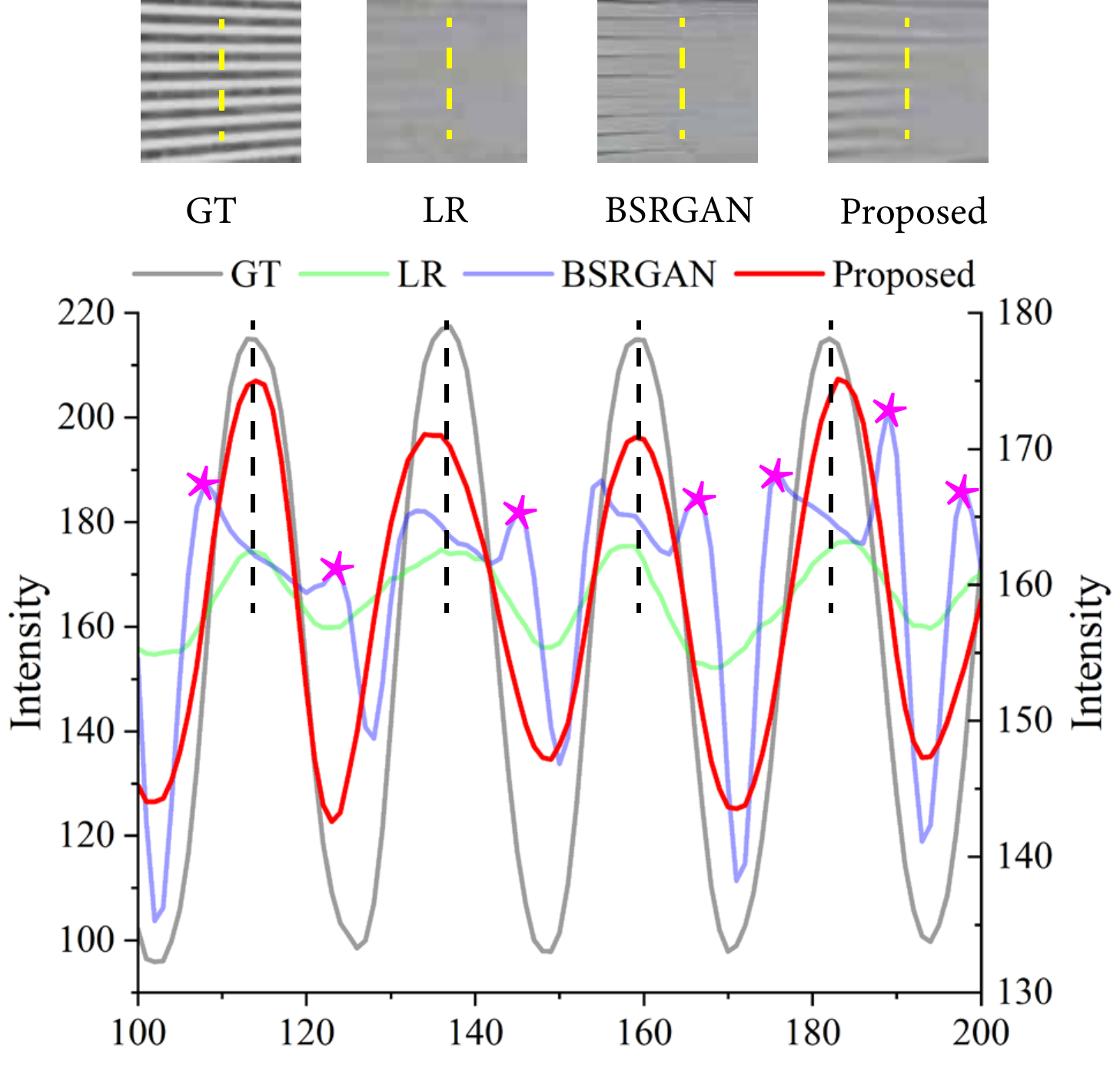}
\vspace{-5pt}
\caption{{Comparison the frequency fidelity of SR results on yellow line (location in the third row of Fig. \ref{fig:RealSR1}), including GT, LR, BSRGAN, and Proposed.} ``*'' denotes artifacts in BSRGAN.}
\label{fig:Results_modu1}
\end{figure}

\NParagraph{Reconstruction fidelity:}
We further investigate the reconstruction fidelity by looking at the profile along the yellow line marked in the top of Fig. \ref{fig:Results_modu1}; GT, LR, BSRGAN, and Proposed are included for the comparison and the results are plotted in the bottom of Fig. \ref{fig:Results_modu1}. 

Despite of the similar visual conception of BSRGAN and Proposed in Fig. \ref{fig:RealSR1}, Fig. \ref{fig:Results_modu1} tells a completely different story. While HR, LR, and Proposed all manifest same periodicity, artifacts are observed from BSRGAN's reconstruction as denoted by the ``*''s. 

\Paragraph{Cross-camera evaluation:}
%
To demonstrate the generalization ability of our proposed model from a source system to a known target system, we also train EDSR by {Ours$^\dag$} datasets (denoted by {Proposed$^\dag$}). As shown Table \ref{tab:tab2}, {Proposed$^\dag$} are comparable to Proposed with a mere 0.1 dB lead in PSNR for Canon testing dataset, and a 0.12 dB and a 0.002 gap in PSNR and SSIM for Nikon testing dataset, respectively. Visually, we can hardly notice the difference between Proposed and Proposed$^\dag$ as shown in Fig. \ref{fig:RealSR1}.

\section{Conclusion}

To address the challenges imposed by the real-world SISR tasks, we propose a frequency-aware physics-inspired degradation model. Unlike the existing approaches to generate synthetic dataset, the proposed model is rooted in Gaussian optics and the physical modelling of camera sensors, which makes it more interpretable and generalizable. Based on the SISR dataset synthesized by our proposed degradation model, we train a deep SISR network for real-world SISR. Experimental results on real-world images have shown that the network trained on our dataset could successfully recover higher spatial-frequency components with less artifacts. In the future, we plan to further refine the degradation-learning process and design a novel SISR algorithm that could full exploit the proposed datasets.

\clearpage

\bibliographystyle{plain}
\bibliography{main}

\newpage

\begin{center}
	\textbf{\large Supplementary Material}
\end{center}

We provide more details and visual results in the supplementary material. Section ~\ref{sec:Sup_Interpretability} provides visualization of real-world SISR degradation mdoel. Algorithm1 and Algorithm2 in the proposed method are described in detail
in Section ~\ref{sec:Algorithm}. More qualitative $\times4$ SR results on real-world SISR datasets and author-captured images are presented in Section ~\ref{sec:Results}.

\section{Interpretability}

\label{sec:Sup_Interpretability}
\begin{figure*}[!t]
\footnotesize
\centering\includegraphics[width=1\linewidth]{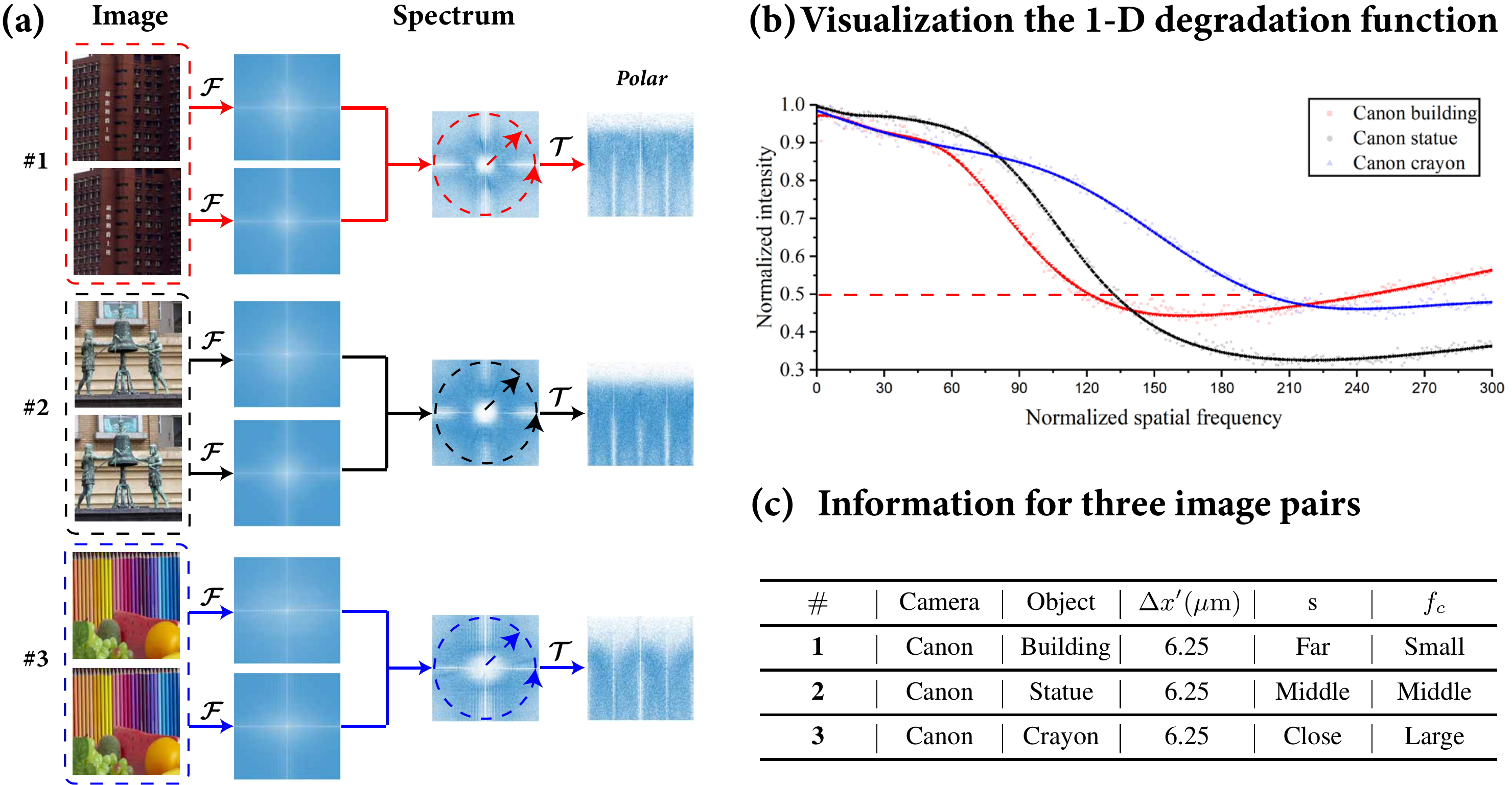}

\caption{{Visualization of real-world degradation model}. (a) A series of operations on three HR-LR pairs of RealSR \cite{cai2019toward} in order to extract real degradation information in the frequency domain. (b) Visualization the 1-D degradation function; (c) Some important information for three image pairs and cameras. $\mathcal{F}$ is 2-D Fourier transform and $\mathcal{T}$ is the conversion of Cartesian coordinate  to polar coordinate.}
\label{fig:sup_degra_func}
\vspace{-10pt}
\end{figure*}

In this section, we further illustrate the dependency of the degradation model on the object  distance $s$, which is shown in Figure \ref{fig:sup_degra_func}. We perform operations on the three image pairs in the Figure \ref{fig:sup_degra_func}(a), and more detailed information about them is provided in Figure \ref{fig:sup_degra_func}(c). 
We could obtain the same conclusion with the paper by  comparing  the  transfer functions $G_s$ for HR-LR pair $\#1$, $\#2$, and $\#3$ as shown in Figure \ref{fig:sup_degra_func}(b).

\section{Algorithms}
\label{sec:Algorithm}
In this section, we provide two detailed algorithms in the method. Algorthm \ref{alg:algorithm1} is how to extract the cutoff frequency $f_c$ of the degradation function $G_s$ and other is how to synthesize LR images from unpaired HR images, which is listed in Algorthm \ref{alg:algorithm2}.
\begin{algorithm*}
\caption{Extract the  cutoff frequency of the degradation function}
$f_c$ : cutoff frequency in the degradation function\;
$J$ : number of patch pairs in one image pair\;
$\mathcal{C(\cdot)}$ : cropping operation\;
$\mathcal{T(\cdot)}$ : the conversion of Cartesian coordinate  to polar coordinate.\;
$\mathcal{F(\cdot)}$ : 2-D Fourier transform\;
$G_s(\cdot)$ : the 2-D transfer function of image pair\;
$g_s(\cdot)$ : the 1-D transfer function of image pair\;
{$\left\{ {I^{HR},I^{LR}} \right\}\leftarrow$ load one SISR image pair\;
$\left\{ {I_{j}^{HR},I_{j}^{LR}} \right\}\leftarrow \mathcal{C}(\left\{ {I^{HR},I^{LR}} \right\})$\;
\ForEach{${j}$ in $1...J$}
{

{$G_s({I_{j}})\leftarrow \mathcal{F}(I_{j}^{LR})/ \mathcal{F}(I_{j}^{HR})$ \;
}
$g_s({I_j})\leftarrow mean(\mathcal{T}\{ G_s({I_j})\}, 2)$\;
${f_{j_c}}\leftarrow {{g_s}}({I_j}) = 0.5$\;
}
}
\textbf{return} ${f_{j_c}}$
\label{alg:algorithm1}
\end{algorithm*}

\begin{algorithm*}
\caption{Synthesize LR images from unpaired HR images}
$f_c$ : cutoff frequency in the degradation function\;

$J$ : number of patch pairs in one HR image\;
$\mathcal{C(\cdot)}$ : cropping operation\;
$\mathcal{F(\cdot)}$ : 2-D Fourier transform\;
$\mathcal{F}^{-1}(\cdot)$ : 2-D inverse Fourier transform\;
$\mathcal{R(\cdot)}$ : predicted network\;
$\mathcal{B(\cdot)}$ : second-order Butterworth low-pass filter\;
$\alpha$ : object sampling frequency ratio\;

{$\left\{I^{\textrm{HR,unpaired}} \right\}\leftarrow$ load one unpaired real world HR\;
$\left\{ I_j^{\textrm{HR,unpaired}} \right\}\leftarrow \mathcal{C}(\left\{ I^{\textrm{HR,unpaired}} \right\})$\;
\ForEach{${j}$ in $1...J$}
{
{${f_{j_c,\textrm{source}}}\leftarrow \mathcal{R}(\mathcal{F}(I_j^{\textrm{HR,unpaired}}))$ \;
}
}
$f_{c,\textrm{source}} =\frac{1}{J}\sum\limits_j {f_{j_c,\textrm{source}}}$\;
$f_{c,\textrm{target}} = \alpha{f_{c,\textrm{source}}}$\;

${I^{\textrm{synLR}}}\leftarrow\{ {\mathcal{F}^{ - 1}}[\mathcal{F}({I^{\textrm{HR,unpaired}}}) \times \mathcal{B}(f_{c,\textrm{target}})]\} { \downarrow _N} + \eta$\;
}
\textbf{return} $I^{\textrm{synLR}} $
\label{alg:algorithm2}
\end{algorithm*}

\section{Qualitative Results}

\label{sec:Results}

\begin{figure*}[t]
\centering\includegraphics[width=1.0\linewidth]{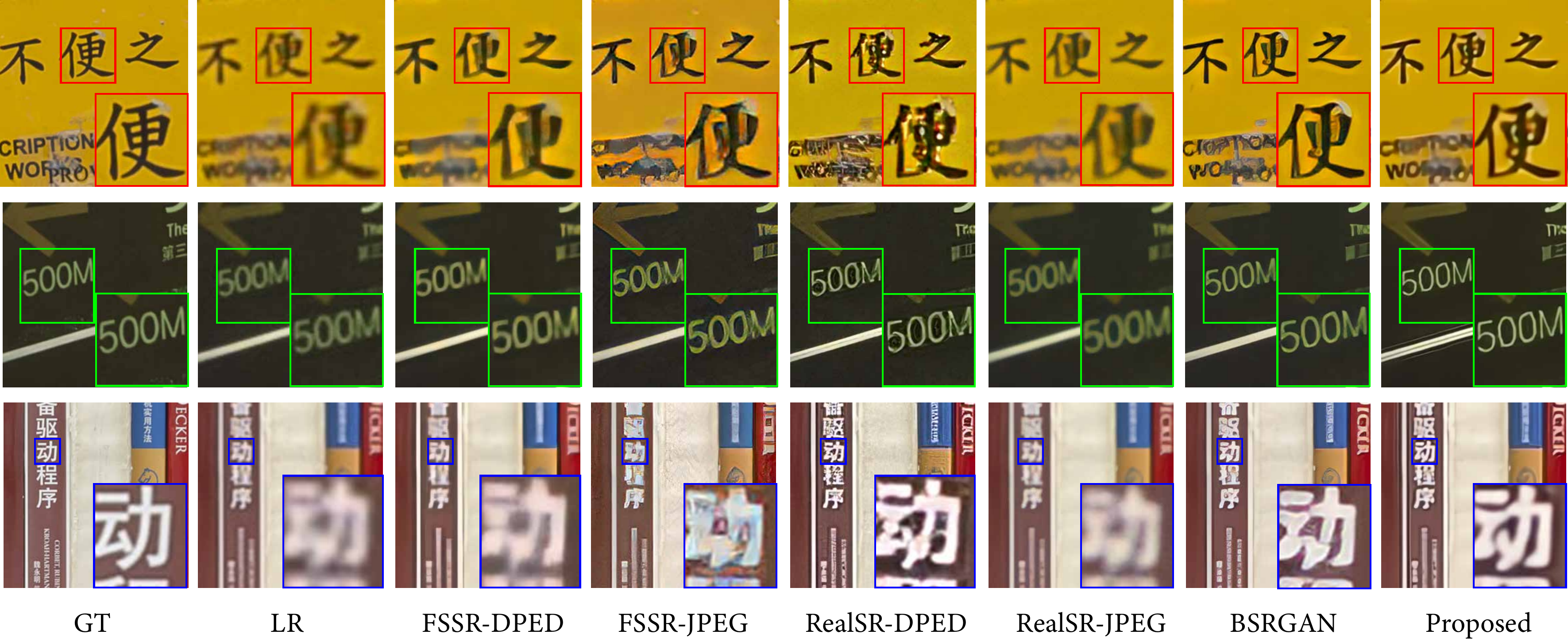}
\vspace{-15pt}
\caption{{Comparison Qualitative $\times4$ SR results of on Canon dataset from RealSR \cite{cai2019toward}, including FSSR-DPED \cite{9022593}, FSSR-JPEG \cite{9022593}, RealSR-DPED  \cite{Ji_2020_CVPR_Workshops}, RealSR-JPEG \cite{Ji_2020_CVPR_Workshops}, and BSRGAN \cite{zhang2021designing}, and Proposed. Our method could recover more natural rendering and less artifacts.}}
\vspace{-5pt}
\label{fig:sup_Canon}
\end{figure*}

\begin{figure*}[t]
\centering\includegraphics[width=1.0\linewidth]{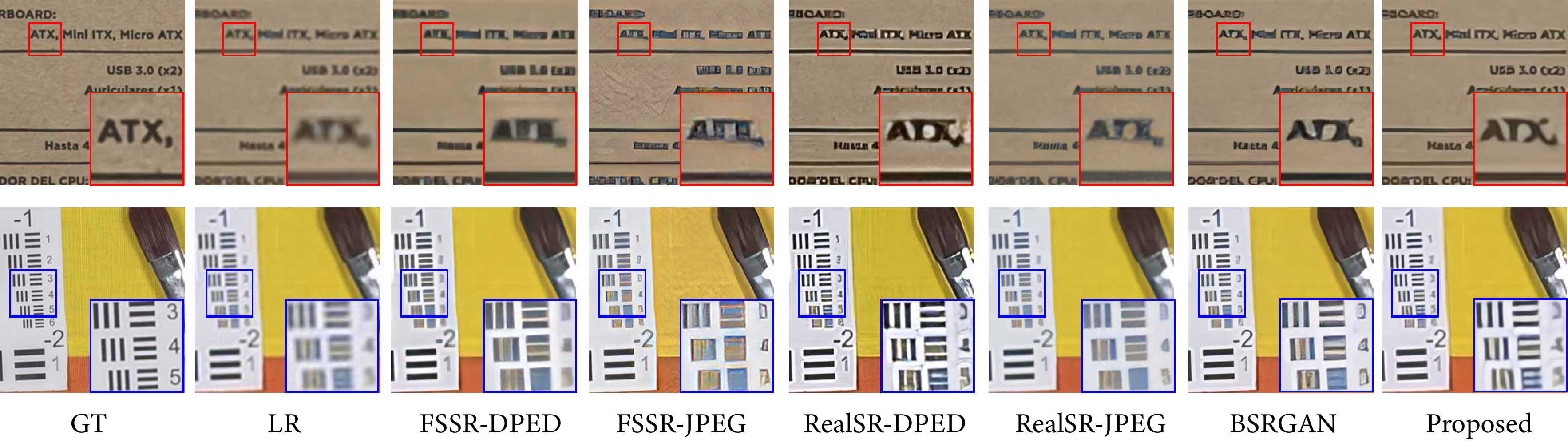}
\vspace{-15pt}
\caption{{{Comparison Qualitative $\times4$ SR results of on Nikon dataset from RealSR \cite{cai2019toward}, including FSSR-DPED \cite{9022593}, FSSR-JPEG \cite{9022593}, RealSR-DPED  \cite{Ji_2020_CVPR_Workshops}, RealSR-JPEG \cite{Ji_2020_CVPR_Workshops}, and BSRGAN \cite{zhang2021designing}, and Proposed. Our method could recover more natural rendering and less artifacts.}}}
\vspace{-5pt}
\label{fig:sup_Nikon}
\end{figure*}

\begin{figure*}[t]
\centering\includegraphics[width=1.0\linewidth]{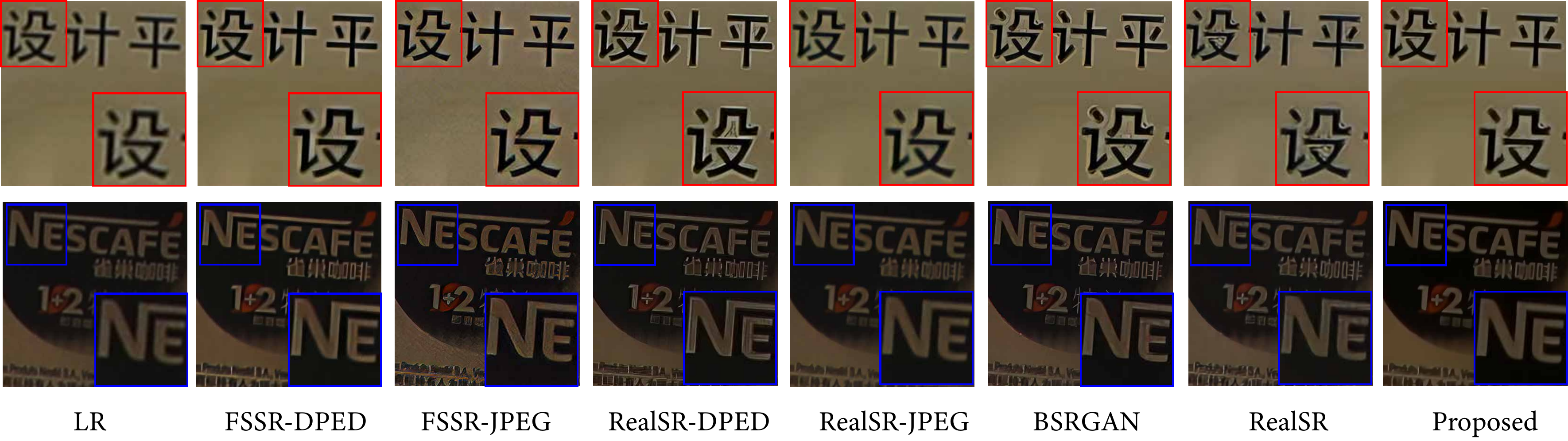}
\vspace{-15pt}
\caption{{Comparison Qualitative $\times4$ SR results of on author-captured images, including FSSR-DPED \cite{9022593}, FSSR-JPEG \cite{9022593}, RealSR-DPED  \cite{Ji_2020_CVPR_Workshops}, RealSR-JPEG \cite{Ji_2020_CVPR_Workshops}, and BSRGAN \cite{zhang2021designing}, EDSR trained on RealSR (denoted by RealSR), and Proposed. Our method could recover more natural rendering and less artifacts.}}
\vspace{-5pt}
\label{fig:sup_Real_image}
\end{figure*}

In this section, we provide additional $\times4$ SR results. A qualitative comparison with SOTAs on the Canon
and Nikon datasets from RealSR \cite{cai2019toward} are shown in Figure \ref{fig:sup_Canon} and Figure \ref{fig:sup_Nikon}, respectively. Figure \ref{fig:sup_Real_image} provides a visual comparison of our proposed method to SOTAs on the author-captured images.

\end{document}